# Transverse Chiral Optical Forces by Chiral Surface Plasmon Polaritons

M.H. Alizadeh[‡,¶] and Björn M. Reinhard[†,¶]

[‡]Department of Physics, [†]Department of Chemistry, and [¶]The Photonics Center,

Boston University, Boston, MA 02215, United States

**Abstract:**

*Recently the concepts of transverse spin angular momentum and Belinfante spin momentum of evanescent waves have drawn considerable attention. Here, we investigate these novel physical properties of electromagnetic fields in the context of chiral surface plasmon polaritons. We demonstrate, both analytically and numerically, that chiral surface plasmon polaritons possess transverse spin angular momentum and Belinfante momentum with rich and non-trivial characteristics. We also show that the transverse spin angular momentum of chiral surface plasmon polaritons leads to the emergence of transverse optical forces in opposite directions for chiral objects of different handedness. The magnitude of this transverse optical spin force on chiral particles is comparable to the magnitude of the forces arising due to intensity gradients and scattering forces. This finding may pave the way for realization of optical separation of chiral biomolecules.*

**Introduction:**

The last few years have seen tremendous interest in novel schemes for optical manipulation of chiral entities.[1-22] This surge is partly fueled by the overall inefficiency of current methods in separation and purification of racemic mixtures of chiral biomolecules.[7, 23, 24] The main idea underlying optical chiral separation is that the interactions of chiral light with chiral molecules depends on their handedness which may lead to their mechanical separation.[2-4, 6, 8] Such a chiral optomechanical separation scheme has been experimentally achieved in microscale using Circular Bragg Reflection.[7] However downsizing this effect to the subwavelength nanoscale, which is the relevant range for the chiral biomolecules, poses a series of significant challenges in overcoming the thermal effects.[25] Several theoretical proposals have been made for accomplishing optomechanical separation of chiral biomolecules. These proposals are mainly based on precise superposition of multiple beams to create local spots with low gradients of electromagnetic (EM) energy density while maintaining large enough gradients of optical chirality.[2, 3, 6] Optical chirality density and the flux of optical chirality fulfill the continuity equation: $\nabla \cdot \boldsymbol{F} + \partial C/\partial t = -1/2(\boldsymbol{j} \cdot \nabla \times \boldsymbol{E} + \boldsymbol{E} \cdot \nabla \times \boldsymbol{j})$, where $\boldsymbol{F} \equiv \{\boldsymbol{E} \times (\nabla \times \boldsymbol{B}) - \boldsymbol{B} \times (\nabla \times \boldsymbol{E})\}/2$ is the optical chirality flux, $C \equiv \frac{\varepsilon_0}{2} \boldsymbol{E} \cdot \nabla \times \boldsymbol{E} + \frac{1}{2\mu_0} \boldsymbol{B} \cdot \nabla \times \boldsymbol{B}$ is the optical chirality density and $\boldsymbol{j}$ is the current density.[26-28] In the presence of a gradient of optical chirality density chiral biomolecules with different handedness will experience forces in opposite directions. Forces of such origin can be practical, if they are comparable in magnitude to the achiral forces in the system. However, due to larger magnitude of the electric polarizability relative to the chiral polarizability, the achiral gradient forces dominate the chiral force. The realization of enantiomer discriminatory forces requires precise control over the amplitudes and especially phases of the interfering beams. Gradients of the optical chirality density can also be achieved in rationally designed plasmonic nano-antennas or metamaterials.[8]. One caveat of this approach is, however, that a plasmonic enhancement of the EM fields boosts both the gradients of the EM fields as well as of the optical chirality. Here we propose a new chiral separation scheme based on transverse Spin Angular Momentum (SAM) of chiral surface plasmon polaritons (SPP). The concept of unusual transverse SAM has recently attracted tremendous attention. [29-34] In the context of SPPs excited by a local chiral source, it was recently shown that they can carry along the chiral character of the source and exhibit non-trivial chiral behavior.[35] Here we extend this idea to include a novel chiral character that chiral SPPs exhibit in terms of chirality-selective chiral optical forces. We demonstrate, both analytically and numerically, that such SPPs possess a distinctive transverse SAM that, in turn, can lead to chiral forces which are in opposite directions for chiral entities of different handedness, including chiral bio-molecules. These chiral forces, due to near-field excitation of the SPPs by a chiral source, have features distinct from those resulting from interference of propagating EM fields. Importantly, they alleviate the need for interference of multiple propagating beams with precise control over their phase and amplitude. Also, the in-plane chiral force is of the same order of magnitude as other forces involved, namely the out-of-plane gradient forces and the in-plane scattering force. We also elaborate on the nature of this chiral transverse force by demonstrating that it arises from the coupling of chiral SPPs and chiral matter through the object's chiral polarizability and we differentiate this force from transverse forces due to Belinfante Spin Momentum, (BSM), which can emerge for achiral objects.[30, 31, 34]

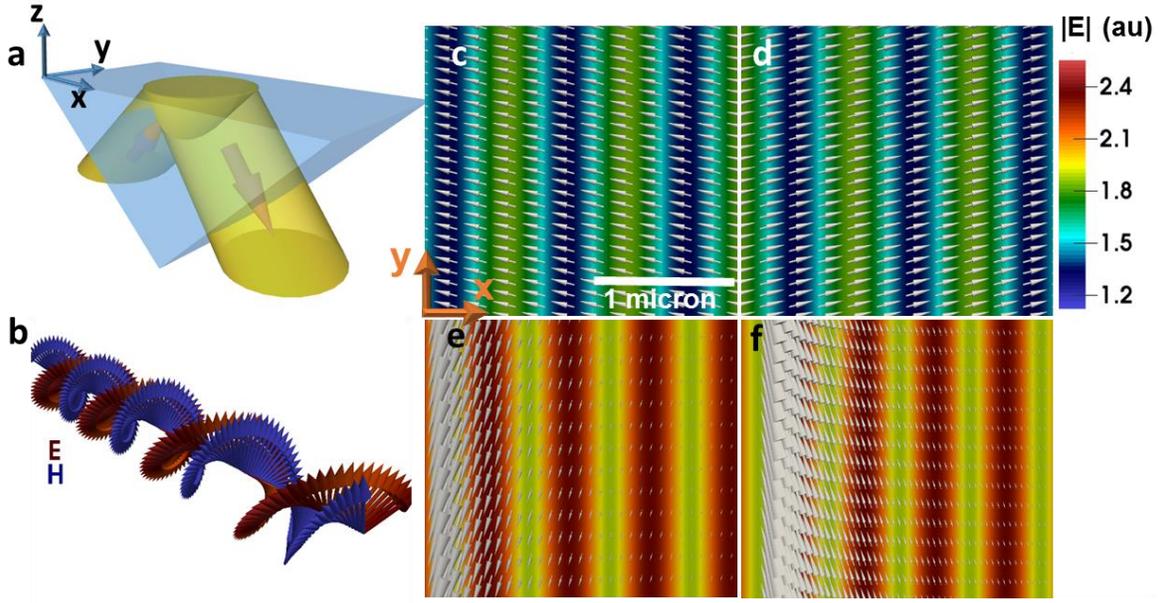

**Figure 1.** (a) Schematics of exciting an evanescent wave in glass-air interface by Total Internal Reflection method. (b) The vectorial behavior of the *E* and *H* of the evanescent wave excited by a circularly polarized beam (c) Distribution of the Spin Angular Momentum (SAM) on XY plane for the resulting evanescent wave near glass surface excited by a right-handed circularly polarized light from the glass side. The background is the map of |*E*|. The coordinates, color and scale bars apply to plots c, d, e and f (d) SAM on XY plane, for the evanescent wave excited by a left-handed circularly polarized light. Despite the expected reversal of the longitudinal SAM, the transverse SAM does not change direction, which shows that the transverse SAM is independent of the polarization of the excitation beam. (e) Belinfante Spin Momentum (BSM) of the evanescent wave excited by a right-handed circularly polarized light. Since its in-plane components are related to the decaying vertical SAM through $B_x \propto \partial_y S_z - \partial_z S_y$, $B_y \propto \partial_z S_x - \partial_x S_z$, and the SAM decays in z direction, a decaying behavior is dictated on $B_x$ and $B_y$. (f) BSM of the evanescent wave on XY plane, excited by a left-handed circularly polarized light.

**SAM and BSM of Chiral SPPs**

The angular momentum of light comprises an orbital and a spin part. Spin angular momentum originates from spinning electric and magnetic fields, while orbital angular momentum stems from the phase gradient of the wavefront.[34, 36-38] This distinction is not fundamental and in fact, spin-to-orbital angular momentum conversion may occur in both inhomogeneous anisotropic media and in tightly focused beams.[39] Spin density is a measure of circular polarization of EM fields and for propagating waves is either parallel or anti-parallel to the direction of the propagation. It can be written as:[31]

$$S = S_e + S_m, S_e = \frac{\varepsilon_0}{4\omega}\text{Im}\{E \times E^*\}, S_m = \frac{\mu_0}{4\omega}\text{Im}\{H \times H^*\} \tag{1}$$

Where $S_e$ and $S_m$ are the electric and magnetic contributions to the SAM. One can readily observe that $S \propto F$. So in the case of SPPs, one can interchangeably speak of SAM or Chirality flux. For evanescent and non-paraxial EM fields, a new type of SAM density can emerge. Recently it was shown that unlike propagating EM fields that can possess SAM only if they are circularly polarized, evanescent waves carry non-zero SAM, which is directed transverse to the wave momentum.[31] For an evanescent wave the transverse SAM originates from the imaginary longitudinal component of the electric field. The longitudinal component induces a rotation of the field in the plane containing both the wave vector and the surface normal and generates a transverse SAM, which is independent of the polarization of the excitation beam. The rotation of the electric field is a consequence of the phase shift between the longitudinal and perpendicular components of the electric field. This phase shift stems from the transversality condition, $k.E = 0$.[32, 33] Changing the polarization of the excitation beam does not alter the direction of rotation of the electric field. It is important to distinguish the transverse SAM from the BSM, defined as $P_b = \nabla \times S$, where $P_b$ denotes the BSM, that originates from the inhomogeneous distribution of the transverse spin, which itself is due to evanescence of the electric and magnetic fields of SPP and leads to non-zero edge current of transverse SAM. New forces and torques in unexpected directions may emerge as a result of transverse SAM and BSM. The key point in distinguishing the origin of the observed forces and torques is to notice that transverse spin is independent of the polarization of the exciting source. Belinfante momentum, however, does depend on the polarization of the source. This can be seen in Figure 1, where the distributions for SAM and BSM of an evanescent field in glass-air interface under circularly polarized excitation are shown. In Figures 1c and 1d, by changing the circular polarization of the excitation beam from left to right, the longitudinal spin of the evanescent wave reverses and as a result its direction changes from parallel to anti-parallel with the wave-vector. The in-plane transverse component of the spin, however, does not change the direction upon polarization inversion of the excitation beam. This is not the case for BSM, which reverses its transverse direction with polarization inversion of the beam (Figure 1e and 1f). Since the excited evanescent waves are propagating in x direction and decaying in z direction, SAM does not decay in the direction of the propagation but decays in z direction. In the case of BSM, since its in-plane components are related to the decaying z component of SAM through $B_x \propto \partial_y S_z - \partial_z S_y, B_y \propto \partial_z S_x - \partial_x S_z$, they decay as the evanescent wave propagates. It should be noted that for a p-polarized evanescent SPP, the situation is starkly different. Due to its strict TM polarization, a p-polarized SPP does not generate any longitudinal SAM and only the transverse component of the SAM survives. (See Figure 2) The vertical decay of the transverse SAM gives rise to a BSM only in the propagation direction with no transverse component. (See Supporting Figure 1). The experimental verification of BSM would require measuring an unusual transverse force on an achiral object. Such an unusual force will be neither in the direction of the wave vector nor aligned with the gradient force, but in a direction dictated by BSM, transverse to the local wave vector of the light. It should be mentioned that using a chiral source with an achiral particle limits the chirality degree of freedom to the EM field only. One major challenge in detecting this novel transverse force is its minuteness compared to the existing gradient and scattering forces. Recently Antognozzi et al. experimentally verified the existence of this unusual transverse force and, thus, demonstrated the presence of BSM in evanescent waves.[30]

This work studied forces exerted on an achiral particle in an evanescent wave which were due to the BSM, which is a polarization-dependent property. So with a change in the polarization state of the incident circular beam, the direction of the transverse force reverses. Here through rigorous analytical and numerical analyses we demonstrate the emergence of novel transverse and chiral-selective forces due to chiral SPPs. We show their unique ability to generate in-plane transverse chiral forces and under appropriate conditions in-plane chiral pulling forces.

**Spin Angular Momentum of Chiral Surface Plasmon Polaritons:**

The propagation of a p-polarized surface plasmon plane wave on a flat surface normal to z direction is described by $\boldsymbol{E} = E(z)\exp(i\boldsymbol{k}\cdot\boldsymbol{\rho} - \omega t)$ where $\omega$ is the angular frequency, $\boldsymbol{k}$ is the in-plane wave vector and $\boldsymbol{\rho}$ is the radial variable in cylindrical coordinates. These surface modes are strictly TM polarized with no transverse electric component. Surface modes excited by a near-field source possess, however, transverse as well as longitudinal components. In the case of a source placed close to metallic interface with the current density of $\boldsymbol{j}(\boldsymbol{r},t)$ the electric field of the SPP can be obtained by: [40, 41]

$$\boldsymbol{E}(\boldsymbol{r},t) = -\mu_0 \int dt' \int d^3 r' \overrightarrow{\boldsymbol{G}'}(\boldsymbol{r},\boldsymbol{r}',t-t') \partial \boldsymbol{j}(\boldsymbol{r},\boldsymbol{r}')/\partial t' \qquad (2)$$

where $\mu_0$ is the vacuum permeability and $\boldsymbol{G}'$ is the Fourier transform of the Green's tensor, $\overleftrightarrow{\boldsymbol{g}}(\boldsymbol{k},z,z',\omega)$, associated with the metal-dielectric infinite interface:

$$\boldsymbol{G}'(\boldsymbol{r},\boldsymbol{r}',t-t') = \int d^2k/4\pi^2 \int d\omega/2\pi \overleftrightarrow{\boldsymbol{g}}(\boldsymbol{k},z,z',\omega) e^{i[k(r-r')-\omega(t-t')]} \qquad (3)$$

It is observed that $\boldsymbol{G}'$ has poles that are determined by the denominators of the Fresnel factors for p-polarized light. These poles relate to the surface plasmon modes and explain the resonant nature of these modes. If the metal slab has finite thickness, plasmon modes can be determined by the zeros of the transcendental equation: $1 + r_{1,2}^{(p)}(k_{sp}) r_{2,3}^{(p)}(k_{sp}) \exp(2ik_{2z}d) = 0$ where $r_{1,2}^{(p)}(k_{sp}), r_{2,3}^{(p)}(k_{sp})$ are the Fresnel reflection coefficients for p-polarized light for upper and lower media respectively. The SPP wave-vector is determined by, $k_{sp} = k_0\sqrt{(\varepsilon_1\varepsilon_2)/(\varepsilon_1+\varepsilon_2)}$ where $k_0$ is the free-space propagation constant and $\varepsilon_1$ and $\varepsilon_2$ are the permittivities of the surrounding medium and metal, respectively.[42, 43] The vertical component of the wave-vector in medium $j$ is determined by $k_{jz} = \sqrt{k_j^2 - k_{sp}^2}$ with $\text{Im}\{k_{jz}\} > 0$. This condition makes the square root single valued. Equation (3) can be used to find the explicit form of the electric field for a generic orientation of dipole placed at a distance d above a flat SPP-sustaining surface. SPPs with chiral character inherit their chirality from the exciting source. Thus, we focus on SPPs excited by an in-plane circularly polarized near field dipole source. Such a source is experimentally feasible to obtain even under normal incident light. One would only need to excite a metal nanoparticle sitting on a metal surface with a circularly polarized light. The light will induce a circularly polarized dipole in the particle and the near field of the particle will have large enough wave vector components to couple to the SPP. This simple scheme lifts the need for phase matching mechanisms through prism-coupling or grating coupling. A schematic drawing of the physical system is shown in Figure 3a, where the near field dipole source can be written as $\mathbf{p} = p_0(\hat{i} + e^{i\pi/2}\hat{j})$, where $\hat{i}, \hat{j}$ are the unit vectors along x and y directions respectively. The resulting SPP is the superposition of the surface plasmons launched normal to each other with a $\pi/2$ phase shift. The resulting circular SPP has a cylindrical symmetry which is evident from the distribution of the z component of the $E$ field, which is shown in Figure 3b. The electric field of a circular SPP is calculated by superposing the electric fields of x-polarized and y-polarized dipoles with a $\pi/2$ phase shift. The explicit form of the electric field can be calculated to be: [35]

$$\boldsymbol{E} = \frac{M(k_{sp},\omega_0)}{\varepsilon_0} p_0 e^{ik_{z1}d} e^{ik_{z1}z} e^{-i\omega t} e^{i\varphi} \{[H_0^{(1)}(k_{sp}\rho) - \frac{1}{k_{sp}\rho} H_1^{(1)}(k_{sp}\rho)]\hat{\rho} - \frac{i}{k_{sp}\rho} H_1^{(1)}(k_{sp}\rho)\hat{\varphi} - \frac{ik_{sp}}{k_{z1}} H_1^{(1)}(k_{sp}\rho)\hat{z}\}$$
(4)

where $M(k_{sp},\omega_0) = \frac{-k_{z1}k_{z2}}{4} \frac{k_{z1}\varepsilon_2 - k_{z2}\varepsilon_1}{\varepsilon_1^2 - \varepsilon_2^2}$, $k_{sp}$ is the in-plane component of SPP wave-vector, $k_{z1}, k_{z2}$ are the vertical components of the wave-vector in air and metal respectively, $\varepsilon_0$ is the vacuum permittivity and $H_0^1(k_{sp}\rho), H_1^1(k_{sp}\rho)$ are different orders of complex Hankel Functions of first kind. The polarization of such an SPP is interesting, as one immediately notices that it has both longitudinal and transverse components. The decaying normal component of the electric field has an intrinsic phase shift relative to the longitudinal component and decays on a length-scale characterized by $k_{sp}\rho$ and $e^{-k_{z1}z}$. This phase shift leads to rotation of the electric field of the chiral SPP in the $\rho z$ plane. This rotating electric field results in a transverse spin in the $\hat{\varphi}$ direction which is transverse to radially outward propagation direction of the SPP. More interestingly, unlike TM polarized SPPs, which have no transverse electric component, there is a transverse electric field component in $\hat{\varphi}$ direction with $-\pi/2$ phase shift relative to the longitudinal component. This unique feature of the chiral SPP bestows it with an in-plane rotation of the electric field and subsequently with a new transverse SAM in z direction. In order to find the analytical expression for the transverse spin of the chiral SPP we make use of Maxwell-Faraday equation to find the magnetic field to be:

$$\boldsymbol{B} = \zeta e^{ik_{z1}z} e^{-i\omega t} e^{i\varphi} \{ -i[\frac{k_{sp}}{k_{z1}} - \frac{1}{k_{sp}\rho}] \frac{H_1^{(1)}(k_{sp}\rho)}{\rho\omega} \hat{\rho} +$$

$$\frac{1}{\omega}[(H_0^{(1)}(k_{sp}\rho) - \frac{1}{k_{sp}\rho} H_1^{(1)}(k_{sp}\rho))k_{z1} + \frac{k_{sp}^2}{k_{z1}} \frac{\partial H_1^{(1)}(k_{sp}\rho)}{\partial(k_{sp}\rho)} \hat{\varphi}] \quad (5)$$

$$+ \frac{2}{\rho\omega}[(H_0^{(1)}(k_{sp}\rho) - \frac{1}{k_{sp}\rho} H_1^{(1)}(k_{sp}\rho))] \hat{z} \}$$

It is evident from equation (5) that the phase shift between the radial and normal as well as the azimuthal components of the magnetic field will produce a rotation of the magnetic field in the $\rho z$ and $\rho\varphi$ planes. Using equations (4) and (5) and plugging into equation (1) one can calculate the electric and magnetic components of the SAM of the chiral SPP to be:

$$\boldsymbol{S}_e = \frac{\varepsilon_0 \omega |\zeta|^2}{2} \{ \Im[\frac{|H_1^{(1)}|^2}{\rho} (\frac{k_{sp}^*}{k_{sp}k_{z1}^*})]\hat{\rho} - \Re[(H_0^{(1)*}(k_{sp}\rho) - \frac{1}{k_{sp}^*\rho} H_1^{(1)*}(k_{sp}\rho))(\frac{k_{sp}}{k_{z1}} H_1^{(1)})]\hat{\varphi}$$

$$- \Re[(H_0^{(1)*}(k_{sp}\rho) - \frac{1}{k_{sp}^*\rho} H_1^{(1)*}(k_{sp}\rho))(\frac{1}{k_{sp}\rho} H_1^{(1)})]\hat{z} \} \quad (6)$$

and

$$\boldsymbol{S}_m = \frac{|\zeta|^2}{\mu_0 \rho \omega} \{ \Im[(\frac{k_{sp}^2}{k_{z1}} + k_{z1})(H_0^{(1)*}(k_{sp}\rho) - \frac{1}{k_{sp}^*\rho} H_1^{(1)*}(k_{sp}\rho))(H_0^{(1)}(k_{sp}\rho) - \frac{1}{k_{sp}\rho} H_1^{(1)}(k_{sp}\rho))]\hat{\rho}$$

$$+ \frac{1}{\rho} \Re[(H_0^{(1)*}(k_{sp}\rho) - \frac{1}{k_{sp}^*\rho} H_1^{(1)*}(k_{sp}\rho))(\frac{k_{sp}}{k_{z1}} - \frac{1}{k_{sp}\rho}) H_1^{(1)}(k_{sp}\rho)]\hat{\varphi} \quad (7)$$

$$- \Re[H_1^{(1)}(k_{sp}\rho)(\frac{k_{sp}}{k_{z1}} - \frac{1}{k_{sp}\rho})(H_0^{(1)*}(k_{sp}\rho) - \frac{1}{k_{sp}^*\rho} H_1^{(1)*}(k_{sp}\rho))(k_{z1}^* + \frac{k_{sp}^{*2}}{k_{z1}^*})]\hat{z} \}$$

where $\zeta = \frac{M(k_{sp}, \omega_0)}{2\varepsilon_0} p_0 e^{ik_{z1}d}$, which contains the decaying factor of the fields in the vertical direction. A more expanded and explicit form of the SAM can be obtained by substituting $k_{sp} = \Re(k_{sp}) + i\Im(k_{sp}), k_{z1} = \Re(k_{z1}) + i\Im(k_{z1})$, where $\Re(k_{sp}), \Im(k_{sp})$ are the real and imaginary parts of the in-plane SPP wave-vector and $\Re(k_{z1}), \Im(k_{z1})$ are the real and imaginary parts of the vertical component of SPP wave-vector. Also, we could substitute $H_{0,1}^{(1)}(k_{sp}\rho) = J_{0,1}^{(1)}(k_{sp}\rho) + iY_{0,1}^{(1)}(k_{sp}\rho)$, where $J_{0,1}^{(1)}(k_{sp}\rho), Y_{0,1}^{(1)}(k_{sp}\rho)$ are cylindrical Bessel functions of the first and second kind, respectively. The permittivity of silver at λ=357 nm is $\varepsilon_r$=-2.145 and $\varepsilon_i$ =0.275, where $\varepsilon_r$ and $\varepsilon_i$ are the real and imaginary parts of the permittivity. Considering that $k_{sp} = k_0\sqrt{(\varepsilon_1\varepsilon_2)/(\varepsilon_1 + \varepsilon_2)}$ and $k_{z1} = \sqrt{k_1^2 - k_{sp}^2}$ , one obtains $\Re k_{sp} = 1.35 k_0$, $\Im k_{sp} = 0.07 k_0$ and $k_{z1} = i 0.914 k_0$ which translate to $\Im k_{sp} \ll \Re k_{sp}$, $\Im k_{sp} \ll \Im k_{z1} \approx |k_{sp}|$ and $\Im k_{sp} \ll |k_{sp}|$. With these specifications and considering the fact that the magnetic contribution to SAM is negligible compared to its electric counterpart, SAM for the chiral SPP can be significantly simplified to

$$S \approx \frac{\varepsilon_0 \omega |\zeta|^2}{2} \{[\frac{(|J_1^{(1)}|^2 + |Y_1^{(1)}|^2)}{\rho \Im k_{z1}}]\hat{\rho} - [(J_1^{(1)}J_0^{(1)} + Y_1^{(1)}Y_0^{(1)})]\hat{\varphi} + [\frac{(J_1^{(1)}J_0^{(1)} + Y_1^{(1)}Y_0^{(1)})}{|k_{sp}|\rho} - \frac{(|J_1^{(1)}|^2 + |Y_1^{(1)}|^2)}{|k_{sp}|^2 \rho^2}]\hat{z}\}$$

(8)

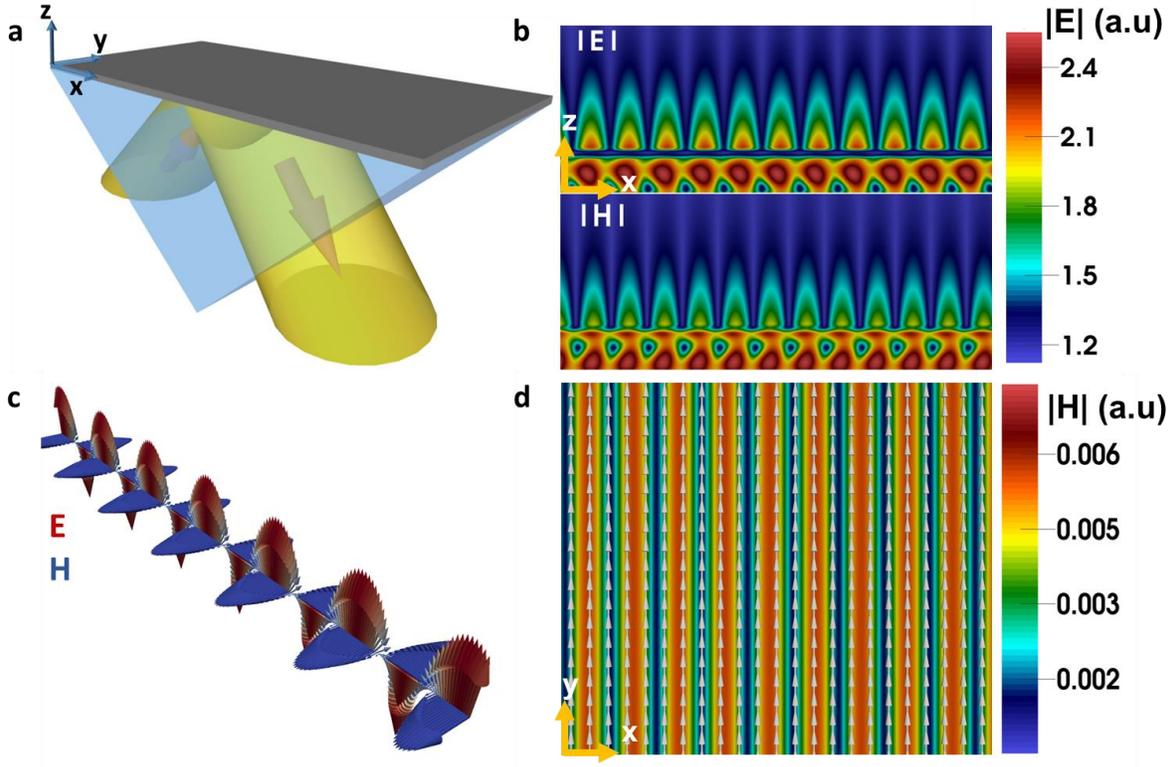

**Figure 2.** (a) Schematics of exciting an SPP from far-field by Kretschmann technique. (b) Maps of |$E$| and |$H$| of the resulting p-polarized SPP are shown on XZ plane. The color bars are in arbitrary units and apply to figures b and d. The excitation of the SPP which is propagating in x direction and evanescent in z direction is evident. (c) Vectorial distribution of the $E$ and $H$ fields of the SPP reveals its TM polarization. (d) SAM of the SPP plotted on XY plane on the background of the $E$ field distribution. The electric field lacks any transverse component which results in zero longitudinal SAM. So unlike the evanescent wave which possesses both longitudinal and transverse components, the SAM of the p-polarized SPP is absolutely transverse.

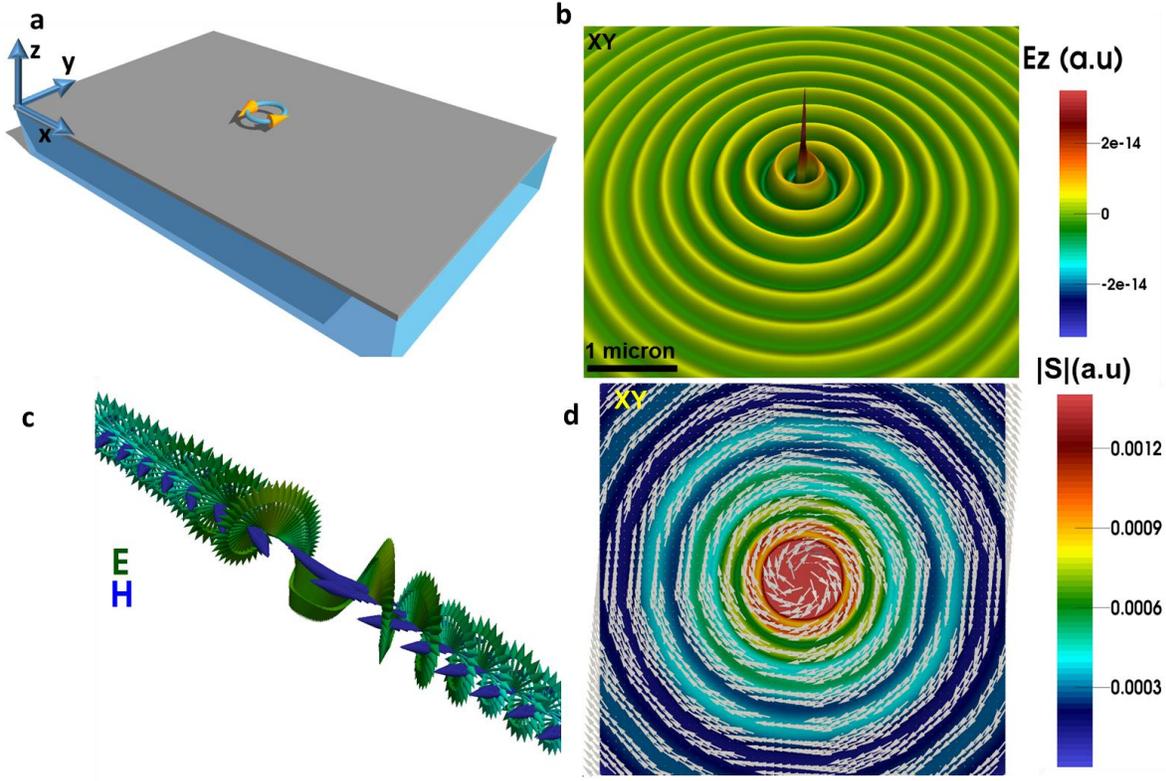

**Figure 3.** (a) Schematics of exciting a chiral SPP by a near-field circularly polarized dipole. (b) The distribution of $E_z$ component of a SPP excited by a near-field circular dipole on XY plane. It is clearly seen that the SPP inherits the circular character of the source. The scale bar applies to figure 3d as well. (c) Vectorial behavior of a chiral SPP, in regions close to the source, is different than that of a p-polarized SPP. For such a SPP *E* and *H* fields possess both longitudinal and transverse components which result in a rich behavior of SAM and optical chirality. In this case the fields are collected along x direction (d) SAM distribution of the SPP excited by a left-handed near-field dipole on XY plane. Close to the origin, the azimuthal transverse SAM is the dominant component. When the polarization of the dipole source reverses, transverse SAM does not change, which indicates its independence from the polarization of the source.

As one would expect, the longitudinal SAM is aligned with the wave vector in $\hat{\rho}$ direction while both in-plane and out-of-plane transverse SAM components are present. it is also noteworthy that the decay length of the in-plane transverse SAM is determined only with that of the cylindrical Bessel functions, while the longitudinal and out-of-plane SAM decay lengths are additionally calibrated with inverse and inverse squared of propagation constant, $k_{sp}\rho$. From equation (8) it is evident that in-plane transverse SAM dominates the spin distribution. This can be seen in Figure 2d, where the SAM for the chiral SPP excited by a left-handed circularly polarized near field dipole source is depicted. The z component of the SAM has been suppressed in these plots. It is evident that the SAM is dominated by the transverse component along $\hat{\varphi}$. When the polarization of the source reverses, transverse SAM does not change its direction. (See Supporting Information)

**Unusual transverse Forces due to Belinfante Spin Momentum**

Belinfante spin momentum, BSM, is defined as the curl of the SAM. if we analogize the SAM of light with bulk magnetization, BSM is similar to the boundary magnetization current or topological Quantum-Hall current in solid-state systems.[33, 44] This vividly explains why BSM is a virtual property of plane waves and yet exists in evanescent waves. In propagating non-dissipating plane waves the local spin loops cancel each other and no net edge current survives, or in other words the curl of the SAM becomes zero. In evanescent waves, however, due to decaying EM fields this cancellation is not complete and a net curl is inevitable.[31] BSM in TM polarized SPPs has been studied elsewhere.[4, 17, 30, 31] Also the emergence of transverse forces on Mie particles due to BSM has recently been verified.[30] One caveat of this transverse force is that it is vanishingly small for Rayleigh particles. Even for larger Mie particles, it turns out to be orders of magnitude smaller than the gradient force and the force due to linear momentum transfer. According to equation (8), the BSM of a chiral SPP will have components in all coordinate directions including the transverse azimuthal direction. So we expect the chiral SPP to exert a transverse force on an achiral object located on the plane of propagation. Moreover, since BSM reverses direction upon polarization inversion of the near-field source, the anticipated transverse force will be in opposite directions for opposite circular polarizations of the exciting near-field source. This can be seen more clearly when one considers the equation of total optical force exerted on a chiral object with chiral polarizability of $\chi$. This force can be written as: [4-6, 9]

$$\langle \boldsymbol{F} \rangle \propto -\nabla U_f + c_{ext} \langle \boldsymbol{\Pi} \rangle + c_{ext} \boldsymbol{P}_b - \Im\{\chi^*(\alpha + \beta/c^2)\} \boldsymbol{S} \tag{9}$$

where, $U_f$ is the EM energy density plus the density of optical chirality, $\boldsymbol{\Pi}$ is the time-averaged Poynting vector, $\alpha$ is the electric polarizability, $\beta$ is the magnetic polarizability and $\chi^*$ is the complex conjugate of chiral polarizability. In lack of particle chirality, i.e. $\chi=0$, the chiral force term vanishes and an emerging unusual transverse force can only be attributed to the transverse components of the BSM. In order to verify the results of the proposed scheme we did a series of full-wavelength EM simulations. We excited a chiral SPP on a silver slab by a circularly polarized near-field dipole source. The metal slab was chosen to be infinitely thick to avoid Fabry-Perot type modes as well as back-reflected waves. A 50 nm spherical gold nano particle was placed on top of the slab and 500 nm away from the source on positive X axis. The distance was chosen to make sure that the fields are dominated by those of the SPP and not of the source. In the next step, the EM forces were calculated numerically using the Maxwell Stress Tensor. (for the details of the simulation see materials and methods) The results are shown in Figures 4a and 4b. Before plotting the forces, we normalized them to the source power. The dipolar source was chosen so as to provide a power of 1fW to the system. Several interesting features are noticed in the force plots. As expected the peaks of the optical forces occur at two resonances of the system, namely the SPP resonance and the localized plasmon resonance of the metal nanoparticle. $F_z$ which mainly represents the involved gradient forces is downward in SPP resonance and upward in localized plasmon resonance. This is because the gradient force tends to push the particle to the regions with higher EM field intensity. At the SPP resonance the fields are strongest near the surface and decay vertically, hence the downward gradient force. At the localized plasmon resonance frequency, however, the EM fields are stronger in the immediate vicinity above the particle. The fields below the particle are weakened in this case due to charge screening by the metal slab. $F_x$ also behaves as expected. Since $F_x$ is an indicative of scattering force, it is strongest when the SPP is excited and travelling radially outward. At the localized plasmon resonance frequency, where the SPP is not excited, such a force resulting from the linear momentum transfer of SPPs is lacking, as is confirmed in the force plots. In the case of $F_y$, unexpected features appear, which cannot be explained in terms of gradient and scattering forces alone. A counterintuitive transverse force, although much smaller, emerges at the resonance frequency of the SPP and, less distinctly, at the localized plasmon resonance frequency. When the circular polarization of the near field source reverses, $F_x$ and $F_z$ remain unchanged, both in magnitude and in direction. $F_y$, however, reverses its direction, which elucidates the origin of the counterintuitive force as a Belinfante spin momentum transfer. One should recall that upon the helicity inversion of the source, transverse SAM does not change directions, while BSM does. This is clearly illustrated in Figures 4c and 4d, where the BSM for two polarizations of the source are plotted. The EM fields at the resonance frequency of the SPP were collected on a plane 50 nm above the metal slab and SAM and BSM were calculated. The BSM reverses its direction upon the polarization inversion of the source, hence generating momenta in opposite directions, which result in opposite forces. It is noteworthy that the enhanced EM fields due to the plasmonic particle results in enhanced BSM.

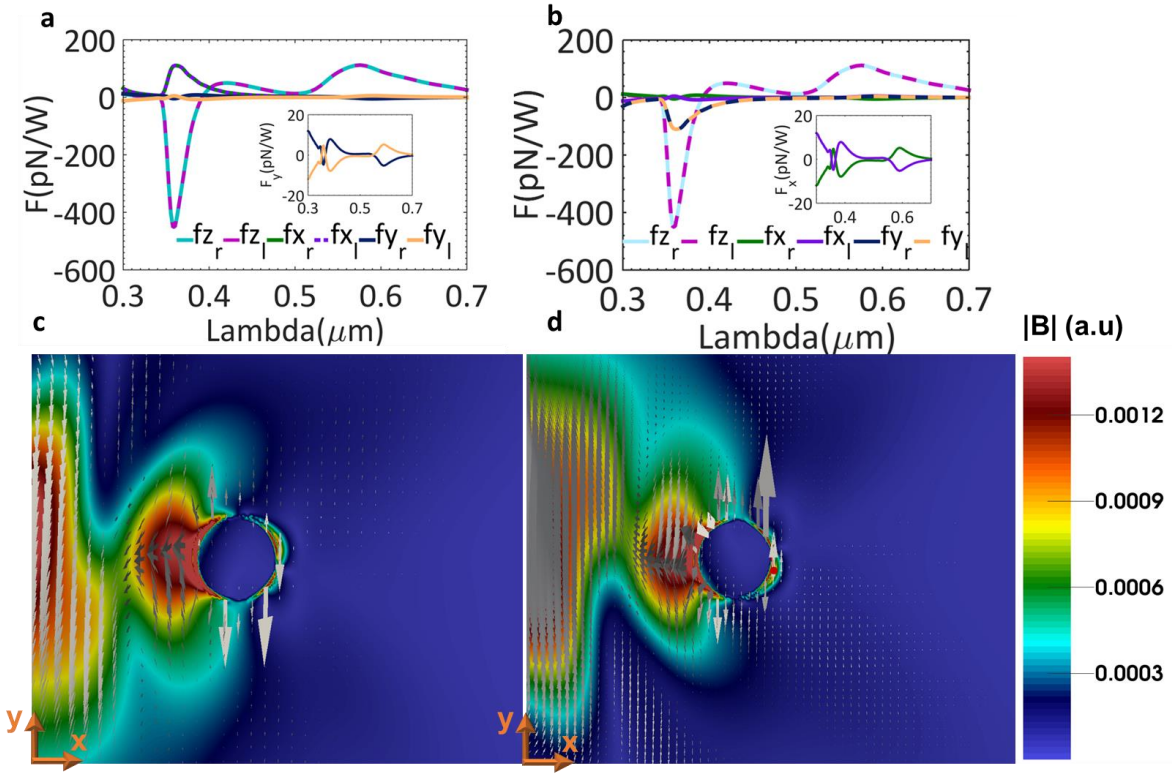

**Figure 4.** (a) The plot of the force exerted by a chiral SPP on a plasmonic nano sphere placed on top of a silver slab on x direction. The appearance of the transverse force in y direction is explained by Belinfante spin momentum. The subscripts r and l denote the right-handed and left-handed polarized dipole sources, respectively. (b) The plot of the force on the same plasmonic nano sphere placed in y direction. (c) Distribution of the BSM for right-handed circular dipole in the vicinity of the particle calculated at the SPP resonance frequency, on a background of the magnitude of the BSM vector, |***B***| (d) Distribution of the BSM for left-handed circular dipole. Unlike the SAM which is independent of the source polarization, the BSM reverses its direction for opposite polarizations of the source.

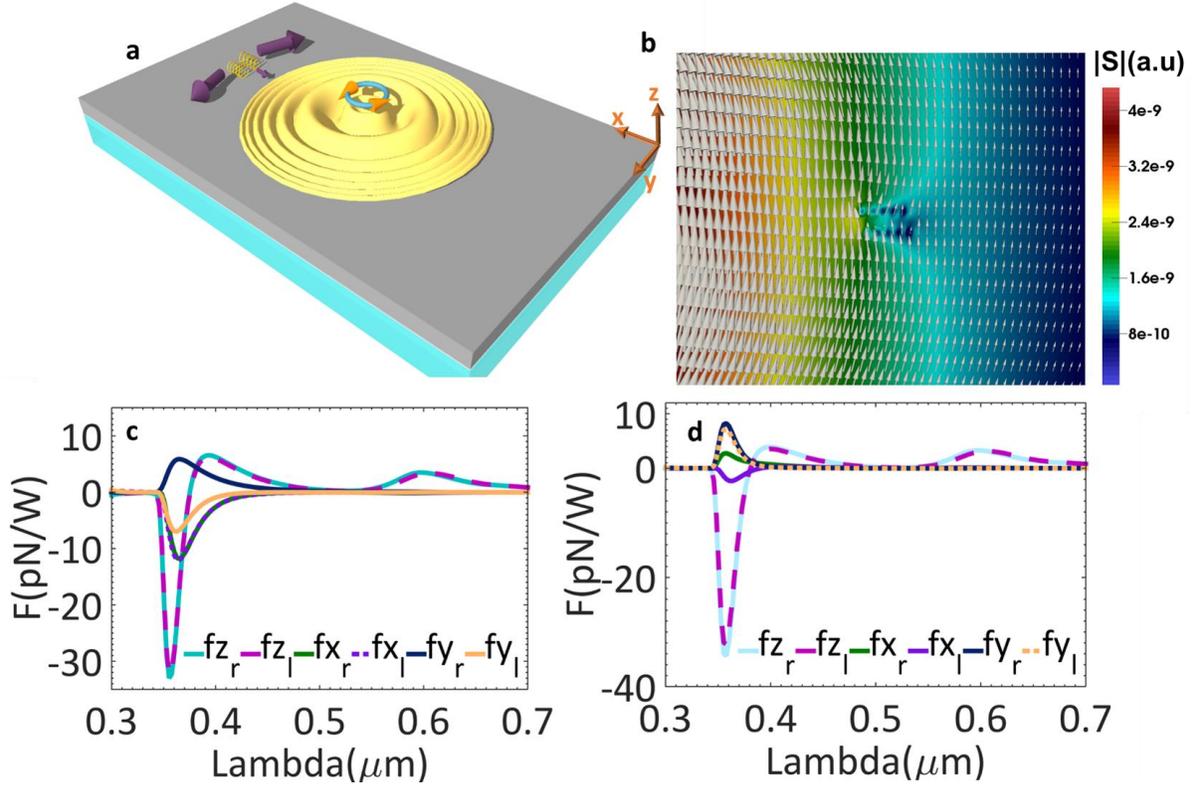

**Figure 5.** (a) Schematics of generating in-plane transverse chiral forces by chiral SPPs. A near-field source excites a chiral SPP, which exerts transverse chiral forces in opposite directions on chiral helices of different handedness upon passing them by. (b) Distribution of the SAM of the chiral SPP, excited by a left-handed circular dipole plotted on a background of the magnitude of the SAM, |*S*|, which is calculated at the SPP resonance frequency. (c) The plot of the forces on two chiral plasmonic particles with different handedness. The particles are placed at positive x direction. The subscripts r and l denote right-handed and left-handed helices, respectively. In the case of chiral particles, it is the transverse SAM which is responsible for the chiral force. (d) The plot of the forces when chiral plasmonic particles are placed at positive y direction.

**Chiral transverse forces due to transverse Spin Angular Momentum**

When cross polarization terms enter the electric and magnetic polarizability of a particle, new force terms emerge which lack for achiral particles. Equation 9 indicates that when the particle is chiral, the SAM of the EM fields can produce optical forces in transverse direction, which are of opposite signs for particles of opposite chirality. Interestingly, the only requirement for the aforementioned transverse chiral forces is the coupling between the transverse component of the SAM and the chiral polarizability of the particle and it occurs for excitation under single polarization of the near-field source. In essence, using a single chiral source, one can generate in-plane transverse forces, which point in opposite directions for objects with opposite $\chi$. A series of Full-Wave EM simulations were carried out to test this hypothesis. We chose a helical metal nanoparticle as a prototypical example of a chiral particle. The dimensions were chosen to mimic a large chiral molecule. The radius of the helix tube was 2.5 nm with helix radius and period of 7.5 nm each and a total number of 5 loops. The chosen size makes our particle ultra-subwavelength. A chiral SPP was excited by a left-handed circularly polarized dipole. As was discussed in detail above, such an SPP has a dominant transverse spin component in azimuthal direction which decays slower than the radial and vertical components. A pair of right-handed and left-handed chiral helices were placed 500 nm away from the source, where the fields from the SPP are dominant. Then the forces were calculated using Maxwell Stress Tensor method. In the first case the particles were placed at x=500 nm, y=0 and z=10 nm. As can be seen in Figures 5c, $F_z$ behaves similarly to the case of an achiral particle. The most prominent feature of $F_z$ is the downward gradient force which occurs at the SPP resonance. The second noticeable gradient force is at the plasmon resonance of the helix at around 610 nm. $F_x$, on the other hand, shows an unexpected behavior. Instead of a scattering force in positive direction, we observe a pulling force toward the source. This pulling force originates from the coupling between the longitudinal SAM and chirality of the particle which has recently been shown for a chiral cluster of plasmonic particles excited by two counter-propagating waves.[9] Interestingly when the helix is placed perpendicular to the longitudinal SAM, under which condition this coupling is zero, this pulling force is eliminated and one observes the usual optical pressure (see Supporting Figure 3). Importantly, even at the localized plasmon resonance frequency no appreciable scattering or pulling force is observed. This is due to lack of any coupling between the chirality of the particle and the SAM, as the SPP is not excited at this wavelength and there is, consequently, virtually zero SAM available. $F_y$, also, exhibits intriguing features under such near-field excitation conditions. For this scenario opposite forces still emerge, albeit with a different origin. In the previous section it was the dipolar source that changed its chirality which resulted in BSM with opposite transverse directions, while the transverse SAM did not change directions. Since the particle was achiral, the force term with coupling between the chirality of the particle and the SAM vanished. In the present case, however, the source does not change its polarization and consequently the BSM maintains its direction, as does the SAM. The handedness of our chiral helices, however, changes which reverses the sign of their chiral polarizability, $\chi$. It is the coupling between the transverse SAM and the chirality of the particle through $\chi$ that results in the observed transverse chiral force. To make sure that this transverse chiral force, follows the pattern of the azimuthal SAM, in the next step, we placed the chiral particles in x=0, y=500 nm and z=10 nm. In this case the main axes of the helices are perpendicular to the longitudinal SAM. For this reason, no pulling forces are observed and the transverse chiral forces occur for $F_x$, as the azimuthal SAM lies strictly in x direction in this specific location. In addition to the observed novel transverse chiral forces, a couple of other points are worth mentioning regarding this chiral force. For one thing, the chiral transverse force has the same order of magnitude as the other forces involved, which is not the case for the transverse forces with the origin of BSM. The relative magnitude should make the chiral force experimentally detectable. Second, one should note that this force directly depends on the chirality factor which is an intrinsic property of the chiral bio-molecules but can be altered for artificial particles. It is also related to the SAM, which is a property of the source. That is why the maximum of the transverse force occurs at the SPP resonance where the transverse SAM reaches its peak due to the SPP resonance and not at the plasmon resonance of the particle.

**Discussion**

In summary, we proposed a new chiral separation scheme based on chiral SPPs. We investigated in detail, the recently proposed BSM and transverse SAM in the context of chiral SPPs. We demonstrated that such SPPs exhibit intriguing and non-trivial properties in terms of transverse SAM and BSM. Unlike p-polarized SPPs, for which SAM is only transverse to the local momentum of the SPP, chiral SPPs possess components of SAM, both aligned with the local momentum as well as transverse to it. The same holds true for the BSM of chiral SPPs. We calculated the forces on a pair of chiral particles with opposite handedness and demonstrated the emergence of enantiomer-selective in-plane chiral forces, transverse to the wave vector of the SPP. This direction is neither aligned with the gradient force, nor with the scattering force. Importantly, for chiral objects of opposite handedness the SPP-mediated forces point in opposite directions. In addition to these unusual transverse forces, an intriguing in-plane pulling force emerges, which is a result of coupling between the chiral dipole moment of the chiral object and the longitudinal SAM. We should emphasize that the more underlying physics of the observed transverse chiral forces is the creation of chiral near-fields which leads to the transverse SAM. Such chiral near-fields can be produced even with a linearly polarized dipolar source, as was recently described in detail.[35] Suffice it to say that such a linearly polarized dipole, e.g. and x-polarized dipole, will launch SSPs only in x direction with vanishing SPP fields as it approaches the y direction. So as long as the chiral particles are placed in the right position, in this case on x axis, they will experience a similar transverse chiral force. Among the possible new areas of optical manipulation that may emerge as a result of these novel forces, optical manipulation of chiral biomolecules is of utmost importance. In this regard, it is important to overcome thermal motion to achieve an optical separation of chiral molecules. Several groups have given estimates of chiral forces on a typical chiral molecule.[4, 6] For instance a hexahelicene, in a frequency range away from its absorption band, has a real chiral polarizability of $\chi \approx 10^{-2}$ $\overset{o}{A}{}^3$, which for an evanescent wave on a glass-air interface, with a beam intensity of 5 mw/μm$^2$, results in a transverse force of about $10^{-4}$ pN, which may be in experimental reach.[4, 6] Chiral SPPs, are expected to yield even larger transverse chiral forces of similar magnitude as optical gradient and scattering forces, potentially paving the path towards novel plasmonic chiral trapping and separation approaches.

**Methods**

**Numerical Simulations**

Full-wave Finite –Difference Time-Domain simulations were used for the simulations. In all the simulations silver was used as the metal medium. The optical constant of silver were extracted from CRC Handbook of Chemistry and Physics, and those of gold were taken from Johnson and Christy.[45, 46] The dipole was positioned at 50 nm above the metal surface and the data were collected from a surface 5nm above the surface. The electric dipole was chosen so that it can emit a power of 1fW. In all the simulations fine meshes with maximum mesh size of λ/70 was used. A fine mesh of λ/350 was used to calculate the Maxwell Stress Tensor.

**Author Information**

Corresponding Authors:

E-mail: halizade@bu.edu (M.H.A.).

E-mail: bmr@bu.edu (B.M.R.).

**Acknowledgement**

This work was supported by the U.S. Department of Energy, Office of Basic Energy Sciences, Division of Materials Science and Engineering under Award DOE DE-SC0010679.

**TOC**

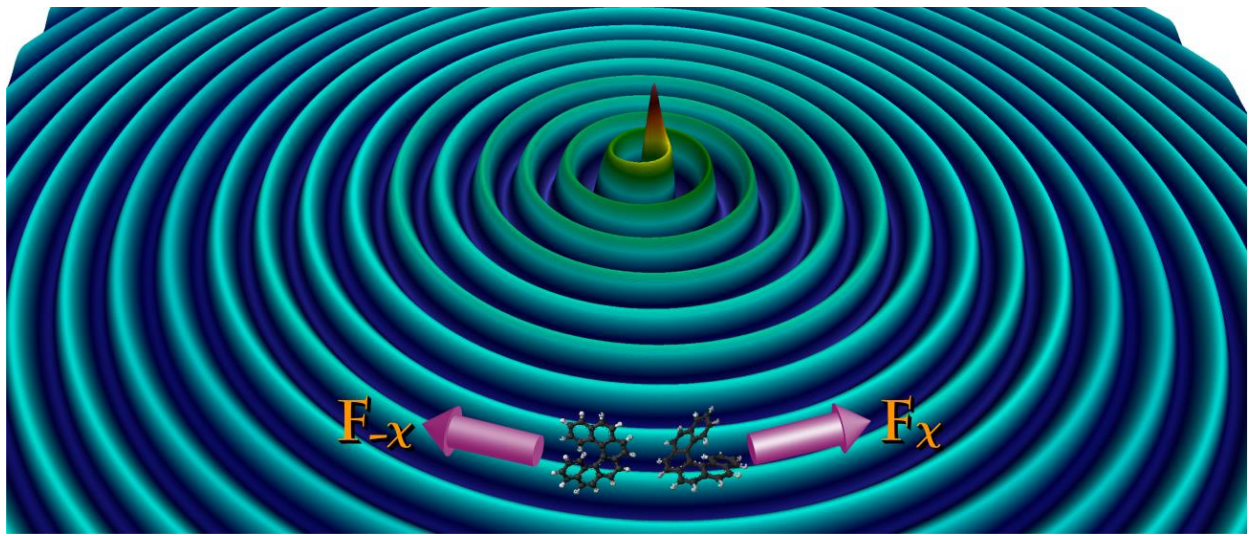